\documentclass[preprintnumbers,amsmath,amssymb]{revtex4}
\usepackage{amsmath}

\usepackage{graphicx}
\usepackage{dcolumn}
\usepackage{bm}

\begin{document}
\title{Solution of one-dimensional Bose Hubbard model in large-$U$ limit}
\author{Yong Zheng}
\email{zhengyongsc@sina.com}
\address{School of Physics and Electronics, Qiannan Normal University for Nationalities,
Duyun 558000, China}
\date{Received \today }

\begin{abstract}
The one-dimensional Bose-Hubbard model in large-$U$ limit has been studied via reducing and mapping the Hamiltonian to a simpler one.  The eigenstates and eigenvalues have been obtained exactly in the subspaces with fixed numbers of single- and double-occupancies but without multiple-occupancies, and the thermodynamic properties of the system have been calculated further. 
These eigenstates and eigenvalues also enable us to develop a new perturbation treatment of the model, with which the ground-state energy has been calculated exactly to first order in  $1/U$.
\end{abstract}

\maketitle

\section{Introduction}
The Bose-Hubbard model perhaps is the simplest one to describe the physics of strongly-correlated Bose systems in a lattice, in which bosons hop between neighboring lattice sites and interact via an on-site repulsion $U$. In general, such model cannot be exactly solved even in the one-dimensional (1D) case \cite{Bt1, Bt2, Bt3}. 

However, in the 1D $U\to \infty$ limit, i.e.,  the hard-core-boson (HCB) case, where double or multiple occupying of a lattice site is prohibited, exact eigenstates of the model can be constructed  easily \cite{HC1,HC2, HC3}. For finite $U$, as in a real system, states with double or multiple occupancies should be considered, and this construction is no longer valid. However, one might expect that,  for the case of finite but large $U$, where the properties of the system are still  dominated by single occupancies or close to that of the HCB case, there should exist some proper ``treatments'' which can simplify the discussion. In this paper, we will focus our discussion on the large-$U$ limit and search for such treatments. 

Another consideration of our discussion comes from perturbative studies of the model. In large-$U$ limit, many previous perturbative studies have taken the on-site-$U$  or potential-energy part of the Hamiltonian as an unperturbed one and viewed the hopping or kinetic part as a perturbation \cite{PT1,PT2,PT3}. These discussions generally  can apply well  only to some special cases, e.g., integer-filling cases. This is mainly due to the highly degenerate eigenstates of the potential-energy part of the Hamiltonian, which lead to many difficulties  in a perturbation calculation. Similar problem has also arisen in the discussion of Fermi-Hubbard model, where it has been  found that a more appropriate treatment of the problem is to retake the unperturbed Hamiltonian by including kinetic terms which do not alter the number of
doubly-occupied sites in a state \cite{FH1,FH2,FH3}. While for Bose-Hubbard model, to the best of our knowledge, similar treatments have not been employed yet. We hope our discussion can go a step further along this line. Actually, a new and simple perturbation treatment will be developed and directly applied to study the ground-state property at the end of our discussion. 

\section{Reduced Hamiltonian and solutions} 

We consider an $N$-particle Bose-Hubbard model on an $L$-site 1D lattice (lattice constant $a=1$), 
\begin{equation}\label{Hb}
	H=-t\sum_{\langle i,j \rangle }c_{i}^{\dagger}c_{j}+U\sum_{i}c_{i}^{\dagger}c_{i}^{\dagger}c_{i}c_{i},
\end{equation}
where $ c_{i}^{\dagger} $ ($ c_{i} $) is the creation  (annihilation) operator of bosons at site $i$, with the periodic boundary condition $ c_{L+1}^{\dagger}= c_{1}^{\dagger}$ ($ c_{L+1}=c_{1} $); $ \langle i,j \rangle $ denotes pairs of nearest-neighbor sites, i.e., $ j=i+1 $ or $ i-1 $, and $t$ is the hopping integral. The on-site $U$ is finite but very large, $U\gg t$, as we has mentioned above.  Here, we also take the restriction $N\le L$, as in the HCB case.  

We first reduce the Hamiltonian into a form which is much easier to solve.  As a proper starting point for treating the large on-site  $U$ beyond HCB approximation, we discuss within the subspace of states which permit each site to be occupied by no more than two bosons.  Then, for our purpose, a site $ i $ can be empty, singly- or doubly-occupied, for which the state can be denoted by  $ |0\rangle_{i} $, $ |1\rangle_{i} $ or $ |2\rangle_{i} $ respectively, with  the on-site constrain $|0\rangle_{i}\langle 0 |+|1\rangle_{i}\langle 1 |+|2\rangle_{i}\langle 2 |=1$.  We can further introduce $ b^{\dagger}_{i}=|1\rangle_{i}\langle 0 | $ ($ b_{i}=|0\rangle_{i}\langle 1 | $) and $ d^{\dagger}_{i}=|2\rangle_{i}\langle 0 | $ ($ d_{i}=|0\rangle_{i}\langle 2 | $), as the creation (annihilation) operators of  single- and double-occupancies respectively at site $i$. Obviously, we have $ b^{\dagger 2}_{i}=b_{i}^{2}=d^{\dagger 2}_{i}=d_{i}^{2}=0 $,  $ d_{i}d_{i}^{\dagger}=b_{i}b_{i}^{\dagger} =|0\rangle_{i}\langle 0 |$, $ [b_{i}, b^{\dagger}_{j}]=\delta_{i,j}(|0\rangle_{i}\langle 0 |-|1\rangle_{i}\langle 1 |)$, $ [d_{i}, d^{\dagger}_{j}]=\delta_{i,j}(|0\rangle_{i}\langle 0 |-|2\rangle_{i}\langle 2 |)$ and $[b_{i}, d_{j}]=0$. Then, within the subspace of states  we discuss, $ c_{i}^{\dagger} =b_{i}^{\dagger}+d_{i}^{\dagger}b_{i} $ and $ c_{i}=b_{i}+b_{i}^{\dagger}d_{i}  $. Replacing $ c_{i}^{\dagger} $'s and $ c_{i} $'s in Eq.~\eqref{Hb} with these relations finally yields a reduced Hamiltonian 
\begin{align*}\label{H}
	H&=H_0+H_I,\\
	H_0&=-t\sum_{\langle i,j \rangle}(b_{i}^{\dagger}b_{j}+b_{i}^{\dagger}d_{i}d_{j}^{\dagger}b_{j})+U\sum_{i}d_{i}^{\dagger}d_{i},\\
	H_I&=-t\sum_{\langle i,j \rangle}\left( d_{i}^{\dagger}b_{i}b_{j}+\text{h.c}\right),  
\end{align*}
where, similar to the case of Fermi-Hubbard model \cite{FH1,FH2,FH3}, we have split the Hamiltonian  into two parts:  the part $H_0$ which preserves the number of singly- or doubly-occupied sites in a state, and the part $H_I$ which would change these numbers. 

Since only the large-$U$ limit is concerned here, the properties of the system are mainly determined by $H_0$, and $H_I$ can be viewed as a perturbation part.

\subsection{Eigenstates and eigenvalues when neglecting $ H_I $} \label{2A}
\begin{figure}
	\centering
	\includegraphics[width=0.8\linewidth]{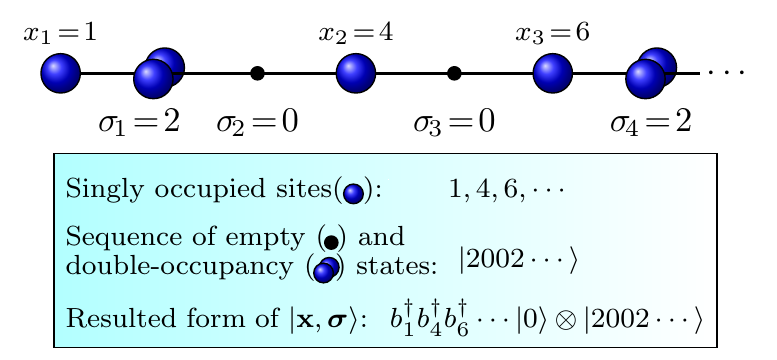}
	\caption{ (Color online) An illustrating example for the notation in $|\mathbf{x},\boldsymbol{\sigma } \rangle$: $x_1,x_2,\cdots$ are the coordinates of singly-occupied sites, and $| \sigma _{1} \sigma _{2} \cdots \rangle$ is the sequence of empty and double-occupancy states along the lattice.}
	\label{fig1}
\end{figure}

Let us first neglect the  perturbation part  $ H_I $, i.e., $ H\approx H_0$ now, which is much easier to solve but remains nontrivial.   
We will find that the eigenstates can be obtained exactly. Consider states with $N_1$ singly-occupied and $N_2$ doubly-occupied sites, i.e., the total number of bosons $N=N_1+2N_2$. Apparently, both  $N_1$ and $N_2$ are conserved by $ H_0 $.
The basis states of the system can be commonly written as $ |\varphi_{\mathbf{x},\boldsymbol{\gamma }} \rangle= \prod_{i=x_1}^{x_{N_1}} b_i^{\dagger }\prod_{j=\gamma _1}^{\gamma _{N_2}} d_j^{\dagger }|0 \rangle $, in which sites $x_1,x_2,\cdots, 
x_{N_1}$ are  singly-occupied and sites $\gamma _{1}, \gamma _{2},\cdots, \gamma _{N_2}$ are doubly-occupied. However, we find that, for our purpose, it is more convenient to rewrite the basis states in another equivalent way. Actually, we note that for each $|\varphi_{\mathbf{x},\boldsymbol{\gamma }} \rangle$, once the singly-occupied sites  $x_1,x_2,\cdots, 
x_{N_1}$ are known, the rest sites on the lattice can only be empty or doubly-occupied; that is, we only need further the arranging information of these empty and double-occupancy states at the rest sites, rather than their specific location information, to completely determine  the state. Then, we can represent the basis states as
\begin{equation}\label{BT}
	|\mathbf{x},\boldsymbol{\sigma } \rangle =|\mathbf{x} \rangle  \otimes  |\boldsymbol{\sigma } \rangle,
\end{equation}
where  $|\mathbf{x} \rangle\equiv\prod_{i=x_1}^{x_{N_1}} b_i^{\dagger }|0 \rangle$ only contains single-occupancies, while
$|\boldsymbol{\sigma } \rangle\equiv | \sigma _{1} \sigma _{2} \cdots \sigma _{L-N_1} \rangle$ is the sequence of  empty and double-occupancy states on the lattice, with $\sigma _{j}=0$ or $2$, denoting empty or double-occupancy respectively. An illustrating example for this form of basis states is shown in Fig.~\ref{fig1}. 
The advantage of this new form of basis states is that it enables us to separately treat the  single-occupancy part of the states. 
Actually, as far as the part $|\mathbf{x} \rangle$ is considered, the Jordan-Wigner transformation is applicable, which maps the operators $b_{i}^{\dagger}$ and $ b_{i} $ to spinless-fermion creation and annihilation operators $f_i^{\dagger }$ and  $f_i$ respectively \cite{HC1,HC2, HC3}:   $b_{i}^{\dagger}\rightleftharpoons f_{i}^{\dagger} \prod_{m=1}^{i-1}(-1)^{\hat{n}^f_{m}}$ and $ b_{i}\rightleftharpoons\prod_{m=1}^{i-1}(-1)^{\hat{n}^f_{m}} f_{i}$, where $ \hat{n}^f_{m}\equiv f_{m}^{\dagger}f_{m} $. Here
we can directly map the part $|\mathbf{x} \rangle$ in Eq.~(\ref{BT}) to $N_1$-particle basis states of spinless fermions $\widetilde{{|\mathbf{x} \rangle} }\equiv\prod_{i=x_1}^{x_{N_1}} f_i^{\dagger }|0 \rangle$, and further map $|\mathbf{x},\boldsymbol{\sigma } \rangle$ to  tensor-product states of $\widetilde{|\mathbf{x} \rangle}$ and $ |\boldsymbol{\sigma } \rangle $, which we can write as
\begin{equation}\label{BT}
	|\mathbf{x} \otimes \boldsymbol{\sigma } \rangle = \widetilde{|\mathbf{x} \rangle} \otimes |\boldsymbol{\sigma } \rangle.
\end{equation}

We can equivalently transform $H_0$ into the space of these states, which we can call the tensor-state space (TSS), noting that $|\mathbf{x} \otimes \boldsymbol{\sigma } \rangle $ and $|\mathbf{x},\boldsymbol{\sigma } \rangle$ have a one-to-one correspondence. For simplicity, we can set $1\leq x_1<\cdots<x_{N_1}\leq L$.
Since the numbers of singly- and doubly-occupied sites both are conserved as far as $H_0$ is considered, we can discuss in the subspace of states  with  fixed $N_1$   and $N_2$. Then, the term $\sum_{i}d_{i}^{\dagger}d_{i}$ in $H_0$  can be simply replaced by $N_2$. 
To transform the remaining terms in $H_0$ into the TSS, we can use a procedure which is very similar to that in Refs.~\cite{89} and \cite{zy}.

Since terms such as $b_{j+1}^{\dagger}b_{j}+b_{j+1}^{\dagger}d_{j+1}d_{j}^{\dagger}b_{j}$ ($j\neq L$) in $H_0$ only transfer a single-occupancy state from site $j$ to $j+1$, without changing the sequence of empty and double-occupancy states on the lattice, that is,  their action on a state $|\mathbf{x},\boldsymbol{\sigma } \rangle$  is equivalent to that of $f_{j+1}^{\dagger} \prod_{m=1}^{j}(-1)^{\hat{n}^f_{m}}\prod_{m'=1}^{j-1}(-1)^{\hat{n}^f_{m'}} f_{j}=f_{j+1}^{\dagger}f_{j}$ on $|\mathbf{x} \otimes \boldsymbol{\sigma } \rangle $. Hence, for $j\neq L$, we can map $b_{j+1}^{\dagger}b_{j}+b_{j+1}^{\dagger}d_{j+1}d_{j}^{\dagger}b_{j}$ to the TSS operator $f_{j+1}^{\dagger}f_{j}$. 

While for $j= L$, $b_{1}^{\dagger}b_{L}+b_{1}^{\dagger}d_{1}d_{L}^{\dagger}b_{L}$ would transfer a single-occupancy state from site $L$ to $1$, and simultaneously transfer an empty or double-occupancy state originally at site $1$ to site $L$. Then, if the original sequence of empty and double-occupancy states on the lattice is 
$ |\sigma _{1} \sigma _{2} \cdots \sigma _{L-N_1} \rangle $, it would be changed to $ |\sigma _{2} \cdots \sigma _{L-N_1} \sigma _{1} \rangle $. Hence, there would be 
a cyclic permutation of this sequence. Additionally, for the single-occupancy part, $ b_{1}^{\dagger}$ and $b_{L} $ can be mapped to  $f_{1}^{\dagger} $ and $\prod_{m=1}^{L-1}(-1)^{\hat{n}^f_{m}} f_{L}=(-1)^{N_1-1}f_{L}$ respectively. Hence, we can map $b_{1}^{\dagger}b_{L}+b_{1}^{\dagger}d_{1}d_{L}^{\dagger}b_{L}$ to the TSS operator $(-1)^{N_1-1}f_{1}^{\dagger}f_{L} P$, and similarly,   map $b_{L}^{\dagger}b_{1}+b_{L}^{\dagger}d_{L}d_{1}^{\dagger}b_{1}$ to $(-1)^{N_1-1}f_{L}^{\dagger}f_{1} P^{-1}$, where $P$ is the cyclic permutation operator of the sequence $ |\boldsymbol{\sigma } \rangle $ and $P^{-1}$ is its inverse, both with their action on the $ |\boldsymbol{\sigma } \rangle $ part of the state, i.e.,  $P|\boldsymbol{\sigma } \rangle= |\sigma _{2} \cdots \sigma _{L-N_1} \sigma _{1} \rangle $ and $P^{-1}|\boldsymbol{\sigma } \rangle= |\sigma _{L-N_1}\sigma _{1} \sigma _{2} \cdots  \rangle $ for $|\boldsymbol{\sigma } \rangle= |\sigma _{1} \sigma _{2} \cdots \sigma _{L-N_1} \rangle $.

Then, $H_0$ can be mapped to an equivalent Hamiltonian in the TSS as
\begin{equation}\label{EH}
	H_0^{\mathrm{equ}}=-t\sum_{j\neq L}(f_{j+1}^{\dagger}f_{j}+\text{h.c})
	+(-1)^{N_1}t(f_{1}^{\dagger}f_{L} P+f_{L}^{\dagger}f_{1} P^{-1})+UN_2,
\end{equation}
which is much easier to solve. From the discussion above, for any two states $|\mathbf{x},\boldsymbol{\sigma } \rangle$ and $|\mathbf{x}',\boldsymbol{\sigma }' \rangle$, one can find the matrix-element relation $\langle \mathbf{x}',\boldsymbol{\sigma }'|H_0|\mathbf{x},\boldsymbol{\sigma } \rangle=\langle \mathbf{x}'\otimes\boldsymbol{\sigma }'|H_0^{\mathrm{equ}}|\mathbf{x} \otimes \boldsymbol{\sigma } \rangle$. 

It should be noted that, similar to the case in Fermi-Hubbard model,  $ H_0^{\mathrm{equ}} $ can also be obtained equivalently by defining a unitary transform operator \cite{zy}
\begin{equation*}
	\mathcal{T}\equiv\sum_{\mathbf{x},\boldsymbol{\sigma}} |\mathbf{x} \otimes\boldsymbol{\sigma} \rangle \langle
	\mathbf{x},\boldsymbol{\sigma}|,
\end{equation*}
which satisfies $\mathcal{T}^{-1}=\mathcal{T}^{\dag}$ and can transform  a state $|\mathbf{x},\boldsymbol{\sigma} \rangle$ to its counterpart in the TSS: $\mathcal{T}|\mathbf{x},\boldsymbol{\sigma} \rangle=|\mathbf{x}\otimes\boldsymbol{\sigma} \rangle $. Then,  noting the matrix-element relation for any two states  $\langle \mathbf{x}',\boldsymbol{\sigma}'|H_0|\mathbf{x},\boldsymbol{\sigma} \rangle=\langle \mathbf{x}'\otimes\boldsymbol{\sigma }'|\mathcal{T} H_0 \mathcal{T}^{\dag}|\mathbf{x} \otimes \boldsymbol{\sigma } \rangle $, we obtain $H_0^{\mathrm{equ}}=\mathcal{T}H_0 \mathcal{T}^{\dag}$. With $\mathcal{T}$, we can represent the mapped form in the TSS for any operators in principle, say, for $H_I$.

To diagonalize $H_0^{\mathrm{equ}}$, let us first introduce the eigenstates of  $P$ and $P^{-1}$, as that in Ref.~\cite{zy}. For any sequence configuration $|\boldsymbol{\sigma}^{s}_1\rangle=|\sigma^{s}_{1}\sigma^{s}_{2}\cdots\sigma^{s}_{L-N_1}\rangle$, we can introduce $|\boldsymbol{\sigma}^{s}_{m+1}\rangle=P^m|\boldsymbol{\sigma}^{s}_1\rangle$, where $m=1,2,\cdots$, till some integer $m_s\leq L-N_1$, for which $|\boldsymbol{\sigma}^{s}_{m_s+1}\rangle=|\boldsymbol{\sigma}^{s}_{1}\rangle$ appears for the first time.
Obviously, $m_{s}$ is directly related to the detailed form of
$|\boldsymbol{\sigma}^s_{1}\rangle$. These configurations form an $m_{s}$-dimensional subspace of sequences, with which we can construct 
$m_{s}$ eigenstates of  $P$ and $P^{-1}$ as follows,
\begin{equation}\label{ks}
	\left|\chi_{k_{s}}\right\rangle=\frac{1}{\sqrt{m_{s}}} \sum_{\nu=1}^{m_{s}} e^{-i \nu k_{s}}\left|\boldsymbol{\sigma}_{\nu}^{s}\right\rangle,
\end{equation}
where $ k_s=2\pi m/m_s $, and $ m=0,1,2,\cdots,m_s-1 $. It can be verified that
$ P\left|\chi_{k_{s}}\right\rangle=e^{i k_{s}}\left|\chi_{k_{s}}\right\rangle $  and  $ P^{-1}\left|\chi_{k_{s}}\right\rangle=e^{-i k_{s}}\left|\chi_{k_{s}}\right\rangle $.

Then,  the eigenstates of  $ H_0^{\mathrm{equ}} $  can be written as
\begin{equation*}
	|\psi\rangle=|\varphi\rangle \otimes\left|\chi_{k_{s}}\right\rangle,
\end{equation*}
where $ |\varphi\rangle  $ is the spinless-fermion part of the eigenfunction. We have
\begin{equation}\label{hei}
	H_0^{\mathrm{equ}}|\psi\rangle=h_f(k_{s})|\varphi\rangle \otimes\left|\chi_{k_{s}}\right\rangle,
\end{equation}
where for convenience, we have introduced $h_f(k_{s})\equiv -t(h+h^{\dagger})+UN_2$, with $h\equiv \sum_{j\neq L} f_{j+1}^{\dagger}f_{j} 
-(-1)^{N_1}e^{i k_{s}}f_{1}^{\dagger}f_{L}$ and $h^{\dagger}\equiv \sum_{j\neq L} f_{j}^{\dagger}f_{j+1} 
-(-1)^{N_1}e^{-i k_{s}}f_{L}^{\dagger}f_{1}$.

$h $ or $h^{\dagger} $  can be easily  diagonalized by considering the case of one spinless fermion. The procedure is just a repeating of that in Ref.~\cite{zy}. Assume $h|\Omega\rangle=\eta|\Omega\rangle$, with $|\Omega\rangle=\sum_{l=1}^{L}a_l f^{\dag}_{l} |0\rangle=\sum_{l=1}^{L}a_l |l\rangle$, where the $a_l$ are coefficients. Since $h^{\dag}h|\Omega\rangle=\sum_{l=1}^{L}a_l h^{\dag}h|l\rangle=|\Omega\rangle=\eta h^{\dag}|\Omega\rangle$, we have $h^{\dag}|\Omega\rangle=\eta^{-1}|\Omega\rangle$. Using $h^L|\Omega\rangle=\sum_{l=1}^{L}a_l h^L|l\rangle=-(-1)^{N_1}e^{i k_{s}}|\Omega\rangle=\eta^L|\Omega\rangle$, we have $\eta^L=-(-1)^{N_1}e^{i k_{s}}$. Then, it follows that  $\eta=e^{iq}$, with $q=k_s/L+(2n+1)\pi/L$ for even $N_1$ and $q=k_s/L+2n\pi/L$ for old $N_1$, 
where $n=0,1,2,\cdots,L-1$; and it should be note that $q\pm 2\pi$ and $q$ are equivalent wave vectors. The corresponding eigenstate  $|\Omega_q\rangle=\frac{1}{\sqrt{L}}\sum_{l=1}^{L}e^{iql} |l\rangle=f^{\dag}_{q}|0\rangle$, where

\[f^{\dag}_{q}=\frac{1}{\sqrt{L}}\sum_{l=1}^L e^{iql}f^{\dag}_{l}.\]

Obviously, $|\Omega_q\rangle$ is also the  eigenstate of the whole part $-t(h+h^{\dagger})$ in $h_f(k_{s})$,  with an  eigenvalue   $\epsilon_{q}=-t(e^{iq}+e^{-iq})=-2t\cos q$. Then, we can write $-t(h+h^{\dagger})=\sum_{q} \epsilon_{q} |\Omega_q\rangle \langle\Omega_q|=\sum_{q} \epsilon_{q} f^{\dag}_{q}f_{q}$, or  
\[  h_f(k_{s})=\sum_{q} \epsilon_{q} f^{\dag}_{q}f_{q}+UN_2, \]
and then Eq.~\eqref{hei} becomes
\begin{equation}\label{Hact}
	H_0^{\mathrm{equ}}|\psi\rangle=\left[  \sum_{q} \epsilon_{q} f^{\dag}_{q}f_{q}+UN_2 \right]  |\varphi\rangle \otimes\left|\chi_{k_{s}}\right\rangle,
\end{equation}
from which, we can take $ |\varphi\rangle=|\mathbf{q}\rangle\equiv f_{q_{1}}^{\dag}f_{q_{2}}^{\dag}\cdots f_{q_{N_1}}^{\dag}|0\rangle$, where $q_{1}, q_{2},\cdots ,q_{N_1}$ are any $N_1$ wave vectors which are different from each other. Then, for given $N_2$, the eigenstate of $H_0^{\mathrm{equ}}$ can be finally  written as
\begin{equation}\label{psiqt}
	|\psi_{N_2;\mathbf{q},k_s} \rangle=|\mathbf{q}\rangle\otimes |\chi_{k_s}\rangle,
\end{equation}
with the  eigenvalue $E_{N_2;\mathbf{q},k_s}=\sum_{v=1}^{N_1}\epsilon_{q_v}+UN_2$. 

It should be noted that for $N_2=0$ (the case without double-occupancies), $|\chi_{k_{s}}\rangle =|0 0 \cdots 0 \rangle$ and $k_{s}=0$. Then, the wave functions take the form  $|\psi_{N_2=0;\mathbf{q}, 0} \rangle=|\mathbf{q}\rangle\otimes |0 0 \cdots 0 \rangle $, which  is completely determined by the spinless-fermion part, and our results  are simply reduced to the HCB ones \cite{HC1,HC2,HC3}.  Hence, our discussion can indeed be viewed as a direct extension of the HCB case by including double-occupancies. For later convenience, we abbreviate the notations  of $|\psi_{N_2=0;\mathbf{q}, 0} \rangle$ and the corresponding $E_{0;\mathbf{q},0}$ by $|\psi_{0;\mathbf{q}} \rangle$ and $E_{0;\mathbf{q}}$ respectively. 

The ground state, which we can denote by $|\psi_{0;\mathbf{q}_0} \rangle$, can be obtained by requiring that $E_{0;\mathbf{q}}=\sum_{v=1}^{N}\epsilon_{q_v}$ takes its minimum:

(i) For old $N$, the $N$ wave vectors of spinless fermions in $ |\mathbf{q}_0\rangle= f_{q_{1}}^{\dag}f_{q_{2}}^{\dag}\cdots f_{q_{N}}^{\dag}|0\rangle $ should respectively take the values $0,\pm \frac{2\pi}{L},\pm \frac{4\pi}{L},\cdots, \pm \frac{(N-1)\pi}{L}$, yielding a total-wave-vector $Q=0$. The ground-state energy $E_{0;\mathbf{q}_0}=-2t\sum_{v=-v_F}^{v_F}\cos \frac{2v\pi}{L}$, where the integer $v_F=\frac{N-1}{2}$.

(ii) While for even $N$, the $N$ wave vectors of spinless fermions should respectively take the values $\pm \frac{\pi}{L},\pm \frac{3\pi}{L},\cdots,$ $ \pm \frac{(N-1)\pi}{L}$, yielding $Q=0$ as well. The ground-state energy $E_{0;\mathbf{q}_0}=-4t\sum_{v=1}^{v_F}\cos \frac{(2v-1)\pi}{L}$, with $v_F=\frac{N}{2}$.

\subsection{Thermodynamics} 
Similar to the case of 1D Fermi-Hubbard model \cite{FH2,FHT}, we can also give a discussion of the thermodynamics of the system basing on the  eigenstates obtained above. It is convenient to discuss with the
grand-canonical partition function $ Z=\sum_N \mathrm{Tr} e^{ -\beta (H_0^{\mathrm{equ}}-\mu N)} $,  with $\beta=\frac{1}{k_\text{B}T}$,  where  $ \mu $, the chemical potential of bosons, has been introduced. The trace here can be calculated  with the eigenstates $|\psi_{N_2;\mathbf{q}, 0} \rangle$. Different from the open-boundary case of 1D Fermi-Hubbard model discussed in Refs.~\cite{FH2,FHT}, one may think that the operators $P$ and $P^{-1}$ in $H_0^{\mathrm{equ}}$, which are directly associated with the periodic-boundary conditions, would lead to trouble in our calculation. However, from Eq.~(\ref{hei}) or (\ref{Hact}), $ \langle \psi_{N_2;\mathbf{q}, k_s} | H_0^{\mathrm{equ}}|\psi_{N_2;\mathbf{q}, k_s} \rangle =\langle \psi_{N_2;\mathbf{q}, k_s} | h_f(k_{s})|\psi_{N_2;\mathbf{q}, k_s} \rangle$, and then $Z$ is reduced to
\begin{align*}
	Z&=\sum_N \mathrm{Tr} e^{ -\beta (H_0^{\mathrm{equ}}-\mu N)} =\sum_{N_1,N_2,k_s} \mathrm{tr} e^{ -\beta \left[ h_f(k_{s})-\mu (N_1+2N_2)\right] } \\
	&=\sum_{N_1,N_2,k_s} e^{-\beta N_2(U-2\mu)}\mathrm{tr} e^ {-\beta \left( \sum_{q} \epsilon_{q} f^{\dag}_{q}f_{q} - \mu N_1 \right) },
\end{align*}
where the new trace ``$\mathrm{tr}$'' is only over eigenstates with the same $N_1$, $N_2$ and $k_s$. This  trace depends on $N_2$ or $k_s$ via the dispersion $\epsilon_{q}$.

Similar to the 1D Fermi-Hubbard case \cite{FH2}, we now focus on the thermodynamic limit, i.e., the limit $L\to \infty $,  $N \to \infty $, for which, the wave-vector $ q $, and hence the dispersion $\epsilon_{q}$,  
tends to be continuous. Then, ``$ \sum_{q} $'' can be replaced by ``$ \left( {L}/{2\pi}\right) \int_{0}^{2\pi} dq$'', and the  trace ``$\mathrm{tr}$'' will become  independent of $k_s$ or $N_2$. The partition function is further reduced as 
\begin{equation}\label{z}
	\begin{split}
		&Z=\sum_{N_1}  \bigg[ \mathrm{tr} e^ {-\beta \left( \sum_{q} \epsilon_{q} f^{\dag}_{q}f_{q} - \mu N_1 \right) } \sum_{N_2,k_s} e^{-\beta N_2(U-2\mu)}\bigg] \\
		=&\sum_{N_1} \bigg[ \mathrm{tr} e^ {-\beta \left( \sum_{q} \epsilon_{q} f^{\dag}_{q}f_{q} - \mu N_1 \right) } \sum_{N_2} C_{L-N_1}^{N_2}e^{-\beta N_2(U-2\mu)}\bigg].
	\end{split}
\end{equation}
The factor $C_{L-N_1}^{N_2}$ is resulted from the sum over $ k_s $, which just gives the total number of sequence configurations $|\boldsymbol{\sigma } \rangle $, for given $N_1$ and $N_2$ \cite{Ex}.
Let $Z_1(N_1)=\mathrm{tr}e^ {-\beta \left( \sum_{q} \epsilon_{q} f^{\dag}_{q}f_{q} - \mu N_1 \right)}$,  which is directly related to the partition function for  a system of free spinless fermions, with $\mu$ playing the role of effective  ``chemical potential ''. Obviously,  in the thermodynamic limit, for a given  $\mu$,  $Z_1(N_1)$  takes its dominated value at the most-probable  particle number $N_1=\bar{N}_1$, where $ \bar{N}_1 $ can be determined via the most-probable distribution of spinless fermions $ f_1(q)=\frac{1}{e^{\beta (\epsilon_{q}-\mu)}+1} $,
\begin{equation*}
	\bar{N}_1=	\sum_{q} \frac{1}{e^{\beta (\epsilon_{q}-\mu)}+1}=\frac{L}{2\pi}\int_{0}^{2\pi} \frac{dq}{e^{\beta (\epsilon_{q}-\mu)}+1}.
\end{equation*}
Hence, similar to the discussion in Ref.~\cite{FH2},  in the thermodynamic limit, we can only keep  the terms with $N_1=\bar{N}_1$ in the sum in Eq.~(\ref{z}),
\begin{equation} \label{zz}
	Z\approx Z_1(\bar{N}_1) Z_2,
\end{equation}
with  a negligible error just as that in replacing the grand-canonical partition function by a canonical one;
where 
\begin{equation*}\label{z1}
	Z_1(\bar{N}_1)=\mathrm{tr}e^ {-\beta \left( \sum_{q} \epsilon_{q} f^{\dag}_{q}f_{q} - \mu \bar{N}_1 \right)},
\end{equation*}
which actually is the canonical partition function for $ \bar{N}_1 $ spinless fermions, and 
\begin{equation*}\label{z2}
	Z_2=\sum_{N_2} C_{L-\bar{N}_1}^{N_2}e^{-\beta N_2(U-2\mu)}.
\end{equation*}

Noting that $ N_2 $ is the number of doubly-occupied sites in the system, one can find that $ Z_2 $ is equivalent to the grand-canonical partition function for a ``system'' of double-occupancies, which have a unique energy level $U$ of $(L-\bar{N}_1)$-fold degeneracy (as the factor $ C_{L-\bar{N}_1}^{N_2} $ indicates) and an effective chemical potential $2\mu$. The distribution function for such system of double-occupancies is  $ f_2=\frac{1}{e^{\beta (U-2\mu)}+1} $.

Eq.~(\ref{zz}) indicates that in the thermodynamic limit, our system can be viewed as a combination of two independent subsystems as far as the thermodynamics is considered: one for $ \bar{N}_1 $ spinless fermions and the other for double-occupancies.

Then, in the thermodynamic limit, the density of singly-occupied sites in the system,
\begin{equation}\label{r1}
	\rho_1=\frac{\bar{N}_1}{L} =\frac{1}{L}\sum_{q} f_1(q)=\frac{1}{2\pi}\int_{0}^{2\pi} \frac{dq}{e^{\beta (\epsilon_{q}-\mu)}+1},
\end{equation}
while the density of doubly-occupied sites,
\begin{equation}\label{r2}
	\rho_2=\frac{\bar{N}_2}{L} =\frac{1}{L} \frac{L-\bar{N}_1}{e^{\beta (U-2\mu)}+1}=\frac{1-\rho_1}{e^{\beta (U-2\mu)}+1},
\end{equation}
and the total particle density of bosons,
\begin{equation}\label{r}
	\rho=\frac{\bar{N}_1+2\bar{N}_2}{L} =\rho_1+2\rho_2.
\end{equation}

\begin{figure}
	\centering
	\includegraphics[width=0.8\linewidth]{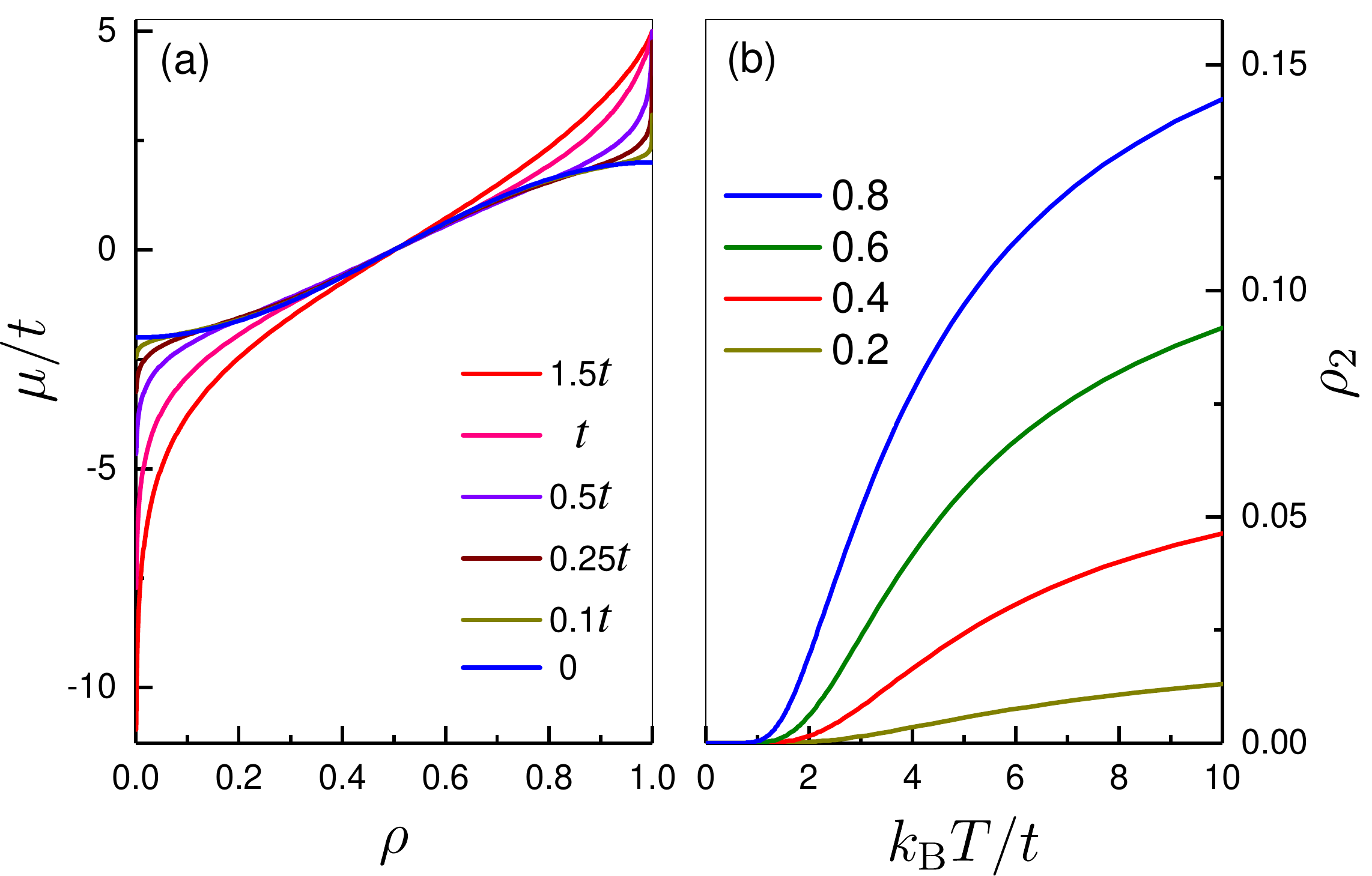}
	\caption{(Color online) (a) Particle-density dependence of $ \mu $ for $k_\text{B}T=0, 0.1t, 0.25t, 0.5t, t, 1.5t$. (b)  Temperature dependence of the density of doubly-occupied sites for $\rho=0.2, 0.4, 0.6, 0.8$.}
	\label{fig2}
\end{figure}

Using Eqs.~(\ref{r1})$-$(\ref{r}), for given particle density $ \rho $, we can calculate  $ \mu $ self-consistently. As an illustration, we take $U=10t$ and show the variation of $ \mu $ with  $ 0<\rho<1$ in Fig.~\ref{fig2}(a) for several temperatures.  
One can find that, there is a gradual departure of the finite-temperature results of $ \mu $ from the zero-temperature one, especially for particle densities $ \rho\sim 0 \text{ or } 1$, reflecting the redistribution of bosons with the increase of temperature. We also show  the temperature-dependence of the density of doubly-occupied sites  $\rho_2$ in Fig.~\ref{fig2}(b), from which we can find that, doubly-occupied sites mainly appear at high temperatures ($k_\text{B}T>t$) and their density is small even at very high temperatures ($k_\text{B}T \sim 10t$), especially for  low-$\rho$ systems. Hence, we can predict that the effect of doubly-occupied sites on the thermodynamic properties of the system is notable only at high temperatures and high particle densities. We can further calculate other interesting thermodynamic quantities of the system in the thermodynamic limit,  such as the internal energy $\bar{E}$,  entropy $S$ and specific heat $C_V$,
\begin{align*}
	&\bar{E}=\frac{L}{2\pi}\int_{0}^{2\pi} \epsilon_{q} f_1(q)dq+(L-\bar{N}_1)Uf_2,\\
	&S=S_1+S_2,\\
	&S_1=-k_{B}\frac{ L}{2\pi}\int_{0}^{2\pi}dq\Big[ f_1(q) \ln f_1(q)  +\left[1-f_1(q)\right] \ln \left[1-f_1(q)\right]\Big], \\
	&S_2= -k_{B}(L-\bar{N}_1)\left[f_2 \ln f_2 +(1-f_2) \ln (1-f_2)\right],\\  	
	&C_V=T\frac{\partial S}{\partial T}.
\end{align*}

\begin{figure}
	\centering
	\includegraphics[width=0.7\linewidth]{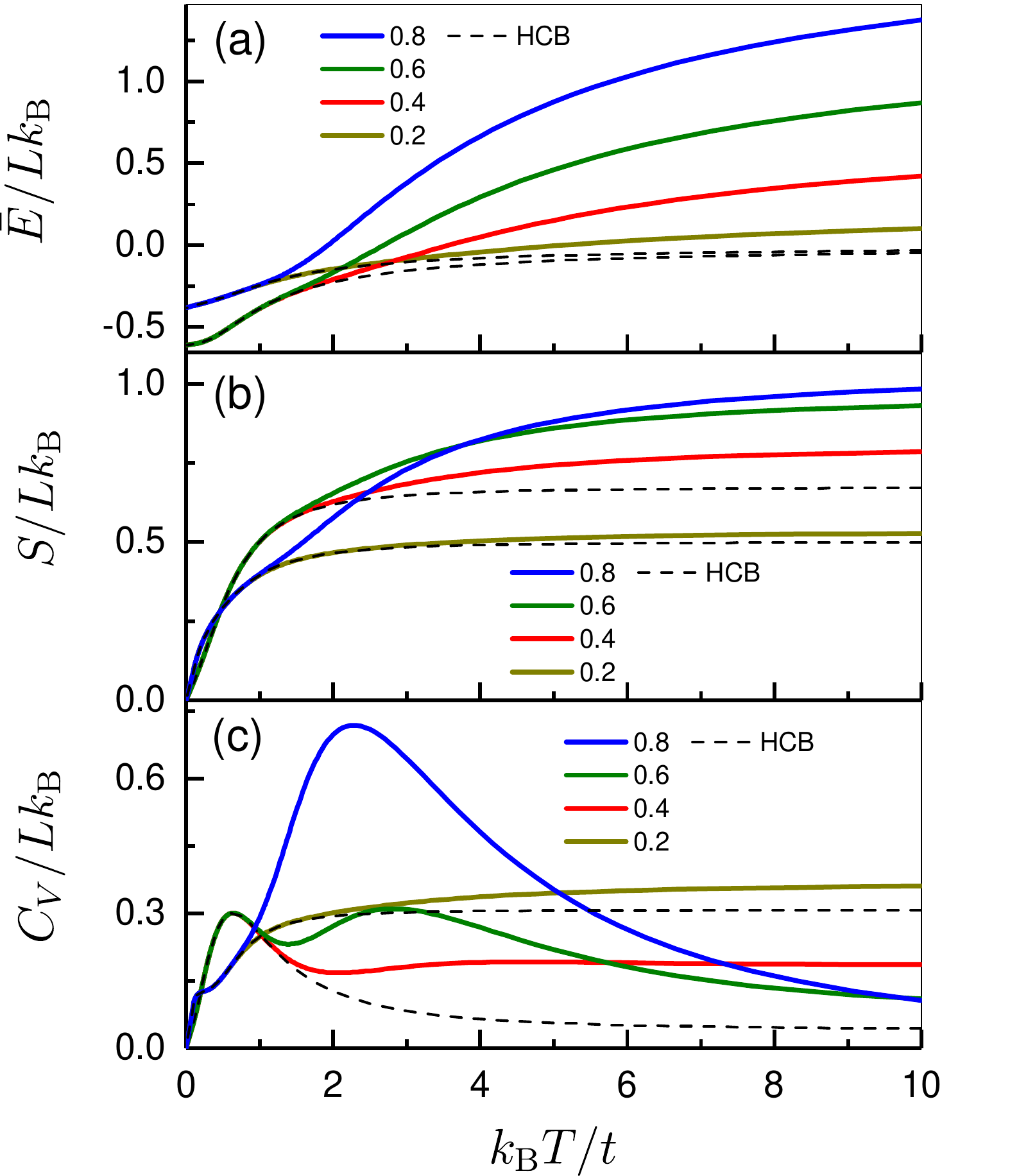}
	\caption{(Color online) Temperature dependence of: (a) internal energy,  (b) entropy and (c) specific heat, for $\rho=0.2, 0.4, 0.6, 0.8$. Also shown are the results of the HCB case (noting that the HCB results are the same for $\rho=0.2$ and $0.8$, or for $\rho=0.4$ and $0.6$). }
	\label{fig3}
\end{figure}

The results are shown in Fig.~\ref{fig3}. For comparison, we have also shown the results of the HCB case, which are obtained by keeping $\rho_1=\rho$.
We can find that, the results of $\bar{E}$, $S$ and $C_V$, all coincide respectively with the corresponding HCB ones at low temperatures ($k_\text{B}T\ll t$). However, at high temperatures ($k_\text{B}T>t$), the departure from the HCB results is obvious, as expected,  since doubly-occupied sites gradually appear in the system with the increase of temperature and  have their effect on the thermodynamic properties of the system mainly at high temperatures.

\subsection{Perturbation treatment of $H_I$}

The eigenstates and eigenvalues obtained above also enable us to  further include $H_I$ to get higher-order approximations or corrections. Since we are most interested in the low-energy cases, especially the ground state, we will mainly discuss how the eigenstates without double-occupancies, i.e., $|\psi_{0;\mathbf{q}} \rangle$,  are corrected by $H_I$. 

To do this, one may perform a canonical transformation $H_s=e^{-iS} H e^{iS}$, with a properly-chosen operator $S$ (see, e.g., Refs.\cite{PT3,FH1,SHS}), to obtain an effective Hamiltonian by ignoring terms which are viewed as high-order ones. However, higher-order terms in the Hamiltonian may not only lead to higher-order corrections, but also can contribute low-order ones to, say, the eigenstates; and hence it is generally hard to see exactly that to what  extent the  eigenvalues and eigenstates of the model have been approximated when using such  treatment. 

Here we want to develop a perturbation treatment, which, as we will see,  is a little different from the usual textbook ones, to study the correction caused by $H_I$. For simplicity, we can first introduce the equivalent operator of $H_I$ in the TSS: $H_I^{\mathrm{equ}}=\mathcal{T}H_I \mathcal{T}^{\dag}$. For any two basis states $|\mathbf{x},\boldsymbol{\sigma } \rangle$ and $|\mathbf{x}',\boldsymbol{\sigma }' \rangle$, we have  
$\langle \mathbf{x}',\boldsymbol{\sigma }'|H_I|\mathbf{x},\boldsymbol{\sigma } \rangle=\langle \mathbf{x}'\otimes\boldsymbol{\sigma }'|H_I^{\mathrm{equ}}|\mathbf{x} \otimes \boldsymbol{\sigma } \rangle$. 
Obviously, $ H_I^{\mathrm{equ}}|\psi_{0;\mathbf{q}} \rangle  $ generates states with $N_2=1$ ( i.e., with one doubly-occupied site). 

To make the discussion not too cumbersome, we will only consider here the correction  to a non-degenerate $|\psi_{0;\mathbf{q}} \rangle$, say, the ground state. We can denote the corrected state by $ |\psi^c_{0;\mathbf{q}} \rangle $ and expand it with the non-corrected eigenstates,
\begin{equation}
	|\psi^c_{0;\mathbf{q}} \rangle =\sum_{\mathbf{q}'}a_{0;\mathbf{q}'}|\psi_{0;\mathbf{q}'}\rangle 
	+\sum_{N_2;\mathbf{q}',k_s}a_{\!N_2;\mathbf{q}'\!,k_s}|\psi_{N_2;\mathbf{q}',k_s} \rangle. \label{SPD}
\end{equation}
The expansion coefficients $ a_{0;\mathbf{q}'}$ and $ a_{\!N_2;\mathbf{q}'\!,k_s} $  can be further expanded in powers of $ {1}/{U} $,
\begin{align}
	&a_{0;\mathbf{q}'}	=a_{0;\mathbf{q}'}^{(0)}+\frac{1}{U} a_{0;\mathbf{q}'}^{(1)}+ \frac{1}{U^2} a_{0;\mathbf{q}'}^{(2)}+\cdots,\label{aq}\\
	&a_{\!N_2;\mathbf{q}'\!,k_s} 	=a_{\!N_2;\mathbf{q}'\!,k_s}^{(0)}+\frac{1}{U} a_{\!N_2;\mathbf{q}'\!,k_s}^{(1)}+ \cdots, \label{am}
\end{align}
where, obviously, $ a_{0;\mathbf{q}'}^{(0)}=\delta_{\mathbf{q}',\mathbf{q}} $ and $ a_{\!N_2;\mathbf{q}'\!,k_s}^{(0)}=0 $. We can also expand the corrected eigenvalue in powers of $ {1}/{U} $,
\begin{equation}\label{Ec}
	E^c_{0;\mathbf{q}}=E_{0;\mathbf{q}}^{(0)}+\frac{1}{U} E_{0;\mathbf{q}}^{(1)}+ \cdots, 
\end{equation}
where $E_{0;\mathbf{q}}^{(0)}=E_{0;\mathbf{q}}$ should be satisfied.

Our perturbation treatment is somewhat different from the usual ones, due to the fact that the small parameter $ {1}/{U} $ is not obviously contained in the perturbation part $H_I$  or $H_I^{\mathrm{equ}}$. Since $ \left( H_0^{\mathrm{equ}}+H_I^{\mathrm{equ}}\right)|\psi^c_{0;\mathbf{q}} \rangle=	E^c_{0;\mathbf{q}}|\psi^c_{0;\mathbf{q}} \rangle  $, we have
\begin{equation*}
	\left( H_0^{\mathrm{equ}}+H_I^{\mathrm{equ}}-E^c_{0;\mathbf{q}}\right)\bigg[ \sum_{\mathbf{q}'}a_{0;\mathbf{q}'}|\psi_{0;\mathbf{q}'}\rangle 
	+\sum_{N_2;\mathbf{q}',k_s}a_{\!N_2;\mathbf{q}'\!,k_s}|\psi_{N_2;\mathbf{q}',k_s} \rangle\bigg]=0.
\end{equation*}
Dotting both sides of this equation with $ \langle \psi_{0;\mathbf{q}'} | $ and $ \langle \psi_{N_2;\mathbf{q}',k_s} | $ respectively yields
\begin{align}
	&\left( E^c_{0;\mathbf{q}}-E_{0;\mathbf{q}'}\right)a_{0;\mathbf{q}'} 
	=\sum_{\mathbf{q}'',k_s}a_{1;\mathbf{q}'',k_s}\langle \psi_{0;\mathbf{q}'} |H_I^{\mathrm{equ}}|\psi_{1;\mathbf{q}'',k_s} \rangle, \label{aqe} \\
	&\left(E^c_{0;\mathbf{q}}- E_{N_2;\mathbf{q}',k_s}\right)a_{\!N_2;\mathbf{q}'\!,k_s}   
	=\sum_{N'_2;\mathbf{q}'',k'_s}a_{\!N'_2;\mathbf{q}''\!,k'_s}\langle \psi_{N_2;\mathbf{q}',k_s} |H_I^{\mathrm{equ}}|\psi_{N'_2;\mathbf{q}'',k'_s} \rangle \notag \\
	&\qquad \qquad\qquad\qquad\qquad\qquad\qquad +\delta_{N_2,1}\sum_{\mathbf{q}''}a_{0;\mathbf{q}''}\langle \psi_{1;\mathbf{q}',k_s} | H_I^{\mathrm{equ}}| \psi_{0;\mathbf{q}''} \rangle. \label{ame0}
\end{align}
For simplicity, we can write $ E_{N_2;\mathbf{q}',k_s}=E_{N_2;\mathbf{q}',k_s}^0+N_2U $, where $ E_{N_2;\mathbf{q}',k_s}^0=E_{N_2;\mathbf{q}',k_s}-N_2U $ obviously is independent of $U$, and then Eq.~\eqref{ame0} becomes
\begin{align}
	&-N_2a_{\!N_2;\mathbf{q}'\!,k_s}+\frac{1}{U}\left( E^c_{0;\mathbf{q}}-E_{N_2;\mathbf{q}',k_s}^0\right)a_{\!N_2;\mathbf{q}'\!,k_s}  \notag   \\
	&=\frac{1}{U}\sum_{N'_2;\mathbf{q}'',k'_s}a_{\!N'_2;\mathbf{q}''\!,k'_s}\langle \psi_{N_2;\mathbf{q}',k_s} |H_I^{\mathrm{equ}}|\psi_{N'_2;\mathbf{q}'',k'_s} \rangle \notag \\
	&\qquad\qquad+\frac{1}{U}\delta_{N_2,1}\sum_{\mathbf{q}''}a_{0;\mathbf{q}''}\langle \psi_{1;\mathbf{q}',k_s} | H_I^{\mathrm{equ}}| \psi_{0;\mathbf{q}''} \rangle. \label{ame}
\end{align}

Substituting Eqs.~\eqref{aq}, \eqref{am} and \eqref{Ec} into Eqs.~\eqref{aqe} and \eqref{ame}, and noting that  matrix elements such as $\langle \psi_{0;\mathbf{q}'} |H_I^{\mathrm{equ}}|\psi_{1;\mathbf{q}'',k_s} \rangle$ and $ \langle \psi_{N_2;\mathbf{q}',k_s} |H_I^{\mathrm{equ}}|\psi_{N'_2;\mathbf{q}'',k'_s} \rangle $ all are independent of $U$,  we collect terms of the same order in $ {1}/{U} $ to obtain:

(i) The zeroth-order terms in $ {1}/{U} $,
\begin{align*}
	\left(E_{0;\mathbf{q}}^{(0)}- E_{0;\mathbf{q}'}\right)a_{0;\mathbf{q}'}^{(0)}&=\sum_{\mathbf{q}'',k_s}a_{1;\mathbf{q}'',k_s}^{(0)}\langle \psi_{0;\mathbf{q}'} |H_I^{\mathrm{equ}}|\psi_{1;\mathbf{q}'',k_s} \rangle, \\
	N_2a_{\!N_2;\mathbf{q}'\!,k_s}^{(0)}&=0,
\end{align*}
from which, we see again that $ a_{\!N_2;\mathbf{q}'\!,k_s}^{(0)}=0 $, $ a_{0;\mathbf{q}'}^{(0)}=\delta_{\mathbf{q}',\mathbf{q}} $ and $E_{0;\mathbf{q}}^{(0)}=E_{0;\mathbf{q}}$.

(ii) The first-order terms in $ {1}/{U} $,
\begin{align*}
	\left(E_{0;\mathbf{q}}- E_{0;\mathbf{q}'}\right)a_{0;\mathbf{q}'}^{(1)}+\delta_{\mathbf{q}',\mathbf{q}}E_{0;\mathbf{q}}^{(1)} 
	&=\sum_{\mathbf{q}'',k_s}a_{1;\mathbf{q}'',k_s}^{(1)}\langle \psi_{0;\mathbf{q}'} |H_I^{\mathrm{equ}}|\psi_{1;\mathbf{q}'',k_s} \rangle, \\
	N_2a_{\!N_2;\mathbf{q}'\!,k_s}^{(1)}&=\delta_{N_2,1}\langle \psi_{1;\mathbf{q}',k_s} | H_I^{\mathrm{equ}}| \psi_{0;\mathbf{q}} \rangle,
\end{align*}
from which, we have $ a_{1;\mathbf{q}'\!,k_s}^{(1)}= \langle \psi_{1;\mathbf{q}',k_s} | H_I^{\mathrm{equ}}| \psi_{0;\mathbf{q}} \rangle$, $ a_{\!N_2 > 1;\mathbf{q}'\!,k_s}^{(1)}= 0$,  $E_{0;\mathbf{q}}^{(1)}=\langle\psi_{0;\mathbf{q}} |\left[ H_I^{\mathrm{equ}}\right] ^2| \psi_{0;\mathbf{q}} \rangle$, and
\[ a_{0;\mathbf{q}'\neq \mathbf{q}}^{(1)}
=\frac{\langle\psi_{0;\mathbf{q}'} |\left[ H_I^{\mathrm{equ}}\right] ^2| \psi_{0;\mathbf{q}} \rangle}{E_{0;\mathbf{q}'}-E_{0;\mathbf{q}}},
\]
where we have used the fact that
\begin{align*}
	&\sum_{\mathbf{q}'',k_s}a_{1;\mathbf{q}'',k_s}^{(1)}\langle \psi_{0;\mathbf{q}'} |H_I^{\mathrm{equ}}|\psi_{1;\mathbf{q}'',k_s} \rangle \\
	=&\sum_{\mathbf{q}'',k_s}\langle \psi_{0;\mathbf{q}'} |H_I^{\mathrm{equ}}|\psi_{1;\mathbf{q}'',k_s} \rangle \langle \psi_{1;\mathbf{q}'',k_s} | H_I^{\mathrm{equ}}| \psi_{0;\mathbf{q}} \rangle  
	=\langle\psi_{0;\mathbf{q}'} |\left[ H_I^{\mathrm{equ}}\right] ^2| \psi_{0;\mathbf{q}} \rangle. 
\end{align*}
While due to the requirement of a normalized $ |\psi^c_{0;\mathbf{q}} \rangle $, $  a_{0;\mathbf{q}}^{(1)} $ can be proved to be a pure imaginary number and can be absorbed as a negligible  phase factor of $| \psi_{0;\mathbf{q}} \rangle$, which is similar to the case in a usual non-degenerate perturbation theory (See, for example, Ref.~\cite{QM}). Hence, we can neglect $  a_{0;\mathbf{q}}^{(1)} $ here.

The results obtained so far can be summarized as follows: 
\begin{align}
	&|\psi^c_{0;\mathbf{q}} \rangle =|\psi_{0;\mathbf{q}}\rangle+\frac{1}{U} H_I^{\mathrm{equ}}| \psi_{0;\mathbf{q}} \rangle \notag\\
	&\quad+\frac{1}{U}\sum_{\mathbf{q}'\neq \mathbf{q}}|\psi_{0;\mathbf{q}'}\rangle \frac{\langle\psi_{0;\mathbf{q}'} |\left[ H_I^{\mathrm{equ}}\right] ^2| \psi_{0;\mathbf{q}} \rangle}{E_{0;\mathbf{q}'}-E_{0;\mathbf{q}}}+O(\frac{1}{U}), \label{Sf} \\
	&E^c_{0;\mathbf{q}}=E_{0;\mathbf{q}}+\frac{1}{U} \langle\psi_{0;\mathbf{q}} |\left[ H_I^{\mathrm{equ}}\right] ^2| \psi_{0;\mathbf{q}} \rangle+ O(\frac{1}{U}). \label{Ecf}
\end{align}
This procedure can continue further to give the detailed form for higher-order terms in principle, but it becomes more and more complicated.

In practical calculations, we always need matrix elements of $(H_I^{\mathrm{equ}})^2$ between states without double-occupancies, as that in Eqs.~\eqref{Sf} and \eqref{Ecf}, which we can calculate as follows:
\begin{align}
	&\langle \psi_{0;\mathbf{q}'} |(H_I^{\mathrm{equ}})^2|\psi_{0;\mathbf{q}}  \rangle= \langle \psi_{0;\mathbf{q}'} |\mathcal{T}H_I^2 \mathcal{T}^{\dag}|\psi_{0;\mathbf{q}} \rangle \notag\\
	=&t^2\langle \psi_{0;\mathbf{q}'} |\mathcal{T}\sum_{\langle i,j \rangle,\langle i,j' \rangle}b_{j'}^{\dagger}b_{i}^{\dagger}d_{i} d_{i}^{\dagger}b_{i}b_{j}\mathcal{T}^{\dag}|\psi_{0;\mathbf{q}} \rangle \notag\\
	=&t^2\langle \psi_{0;\mathbf{q}'} |\mathcal{T}\sum_{i;\delta=\pm 1}\hat{n}_i^b (b_{i+\delta}^{\dagger}b_{i-\delta}+\hat{n}_{i-\delta}^b)\mathcal{T}^{\dag}|\psi_{0;\mathbf{q}} \rangle \notag\\
	=&t^2\langle \psi_{0;\mathbf{q}'} |\bigg[  \sum_{i;\delta=\pm 1}\hat{n}_i^f \hat{n}_{i-\delta}^f-
	\sum_{i\neq L;\delta=\pm 1}\hat{n}_i^f f_{i+\delta}^{\dagger}f_{i-\delta} \notag\\
	&\qquad+(-1)^{N-2}\hat{n}_{L}^f\big[ f_{1}^{\dagger}f_{L-1}+f_{L-1}^{\dagger}f_{1}\big] \bigg] |\psi_{0;\mathbf{q}} \rangle, \label{Heq2}
\end{align}
where $\hat{n}_i^b\equiv b_{i}^{\dagger}b_{i}$, and the last two steps follow through the relation $ d_{i}d_{i}^{\dagger}=b_{i}b_{i}^{\dagger} $ and the  Jordan-Wigner transformation respectively. 

We can take the ground-state correction as an example. Consider the odd-$N$ case (the even-$N$ case can be discussed similarly), of which the non-corrected ground-state result has been discussed in Sec.~\ref{2A}.  

According to Eqs.~\eqref{Ecf} and \eqref{Heq2}, the correction to ground-state energy to first order in  $1/U$ can be calculated as
\begin{equation*}
	\begin{split}
		&\Delta E_{0;\mathbf{q}_0}= \frac{1}{U} \langle \psi_{0;\mathbf{q}_0} | (H_I^{\mathrm{equ}})^2|\psi_{0;\mathbf{q}_0} \rangle +O(\frac{1}{U})\\
		=&\frac{t^2}{U}\langle \psi_{0;\mathbf{q}_0}  |  \sum_{i;\delta=\pm 1}\big[ \hat{n}_i^f \hat{n}_{i-\delta}^f-
		\hat{n}_i^f f_{i+\delta}^{\dagger}f_{i-\delta}\big]  |\psi_{0;\mathbf{q}_0}  \rangle +O(\frac{1}{U})\\
		=&\frac{2t^2}{UL}\sum_{q_1+q_3=q_2+q_4} \left[ \cos (q_2-q_1)  -\cos (q_2+q_1)\right] \langle \psi_{0;\mathbf{q}}|f_{q_4}^{\dagger}f_{q_3} f_{q_2}^{\dagger}f_{q_1}|\psi_{0;\mathbf{q}}\rangle +O(\frac{1}{U}) \\	
		=&\frac{4Nt^2}{UL} \sum_{v=-v_F}^{v_F}\sin^2 \frac{2v\pi}{L}+O(\frac{1}{U}).	
	\end{split}	
\end{equation*}
One can find that $\Delta E_{0;\mathbf{q}_0}\sim \frac{4N^2t^2}{UL}= \frac{4N\rho t^2}{U}$ and the average correction per particle $\Delta E_{0;\mathbf{q}_0}/N\sim \frac{4\rho t^2}{U}$, which can be ignored for large $U$.  

The correction for other non-degenerate $ |\psi_{0;\mathbf{q}} \rangle $ can also be calculated similarly, but with much more complexity, due to the complicated form of Eq.~\eqref{Heq2}. 
The direct extension of our procedure to the case of degenerate $ |\psi_{0;\mathbf{q}} \rangle $ can also be discussed, although it is too lengthy to be presented here.

\section{Conclusion}
In conclusion, our study of the 1D Bose-Hubbard model in the large-$U$ limit is a direct extension of the HCB approximation by including doubly-occupied states. The main part of our reduced Hamiltonian, $H_0$, which perseveres the number of singly- or doubly-occupied sites in a state, enables us to solve it exactly in the TSS we introduced.
With the obtained eigenstates and eigenvalues, we have calculated  the thermodynamic properties of the system. Our results show that double-occupancies mainly appear and affect the properties of the system at high temperatures. 

We think our treatment can capture the main physics of our large-$U$ system. Even though,  a new perturbation treatment has also been developed to discuss the corrections caused by the $H_I$ part, which indeed can be ignored as far as the ground state is considered.  
More further discussions, including the extension of our study to other 1D and quasi-1D Bose systems with large on-site $U$, will be given in future studies. 

\begin{acknowledgments}
	This research was financially supported  by Guizhou Provincial Education Department
	(Grant No.~QJHKY[2016]314) and Qiannan Normal University for Nationalities.
\end{acknowledgments}


\begin{thebibliography}{99}
\bibitem{Ex} Similar factors also appear in the case of 1D Fermi-Hubbard model, even in the open-boundary case, which, however, have been omitted mistakenly in Ref.~\cite{FH2}.
\end{thebibliography}
\end{document}